\documentstyle[12pt]{article}
\newcounter{subeq}

\begin{document}
\begin{titlepage}
\begin{flushright}
OCU-PHYS-154 \\
April, 1994 \\
\end{flushright}
\vfill
\begin{center}
{\Large {\bf Is \lq\lq Heavy Quark Damping Rate Puzzle'' in Hot QCD
Really the Puzzle?
 } }\\
\vfill
{\large {\sc A. Ni\'{e}gawa}\footnote{
E-mail: h1143@ocugw.cc.osaka-cu.ac.jp
}
\vfill
{\normalsize\em Department of Physics, Osaka City University } \\
  {\normalsize\em Sumiyoshi-ku, Osaka 558, Japan} } \\
\vfill
\end{center}
\begin{quotation}
Within the framework of perturbative resummation scheme of Pisarski
and Braaten, the decay- or damping-rate of a moving heavy quark
(muon) to leading order in weak coupling in hot QCD (QED) is
examined. Although, as is well known, the conventionally-defined
damping rate diverges logarithmically at the infrared limit, shown
is that no such divergence appears in the physically measurable
decay rate. The cancellation occurs between the contribution from
the \lq\lq real'' decay diagram and the contribution from the
diagrams with \lq\lq thermal radiative correction''.
\end{quotation}
\vfill
\end{titlepage}

In hot gauge theories, a consistent perturbation scheme (the
hard-thermal-loop (HTL) resummation scheme) has been established
\cite{pis1,bra}. This scheme enables consistent evaluations of any
thermal reaction rates to leading order in the coupling constant.
The HTL resummed propagators screen the infrared singularities, and
render otherwise divergent physical quantities finite, if they are
not sensitive to a further resummation of the corrections of
$O(g^2 T)$. Much interest has been taken \cite{pis1,leb,bur} in the
so-called damping rate of a particle moving in a hot quark-gluon
(electron-photon) plasma, because this is an example of those which
are sensitive to the further resummation mentioned above. In other
words, the damping rate of a moving particle is still divergent
even within the HTL resummation scheme. It is anticipated
\cite{pis1,bur} that the further resummation makes the damping rate
finite.

In this Letter, within the HTL resummation scheme \cite{pis1,bra},
we analyze the decay- or damping-rate of a moving heavy quark
(muon) in a hot quark-gluon (electron-photon) plasma, in terms of a
measurable quantity. We shall find that, although the damping rate
as defined in  a conventional way diverges, no such divergence
occurs in the physically measurable decay rate. The mechanism of
resulting the finite decay rate is the same as the familiar
Bloch-Nordsieck mechanism, which operates to cancel the infrared
divergences in reaction rates in vacuum theories \cite{mut}.

We consider the heavy quark with velocity ${\bf v} =
{\bf p}/E$ ($O(g) << v$) injected into the quark-gluon plasma. By
the term \lq\lq heavy quark'' we mean so heavy a quark that it is
not thermalized, i.e., $e^{-E/T} << 1$. The heavy muon immersed in
the hot electron-photon plasma may be treated in a parallel manner
(see below). We employ the real-time formalism of thermal field
theory \cite{lan}, which is formulated on the time path
$- \infty \to + \infty \to - \infty \to - \infty - i \, T^{-1}$, in
a complex time plane.

The decay rate of the heavy quark with any one of the color states
is obtained as \cite{kob,nie}
\begin{equation}
{\cal R}_d  \equiv
      - i \, \frac{M}{E} \, \frac{1}{2}
      \sum_s \bar{u}_s (P) \, \Sigma_{21} (P) \, u_s (P) \, ,
      \end{equation}
where $\Sigma_{21} (P)$ with $P = (E, {\bf p})$ is the
(21)-component of the self-energy matrix
$\Sigma_{j i} (P) \, \, (i, j = 1, 2)$ of the heavy quark. Here $i$
and $j$ designate the type of heavy quark-gluon vertex
\cite{lan,nie}, the type-1 (type-2) vertex stands for the vertex of
physical or type-1 (thermal-ghost or type-2) fields. [The
\lq\lq Feynman rules'' in the above-mentioned real-time formalism
is equivalent to the circled diagram rules of Kobes and Semenoff
\cite{kob}, provided that the physical (thermal-ghost) field is
identified with the field of \lq\lq uncircled'' (\lq\lq circled'')
type in \cite{kob}.] To one-loop order in the HTL resummation
scheme, $\Sigma_{2 1}$ is diagrammed in Fig. 1 with $i = 1$ and
$j = 2$;
\begin{eqnarray}
\Sigma_{21} (P) =
      - i \, g^2  C_f
      \int \frac{d^{\, 4} Q}{(2 \pi)^4} \,
      \displaystyle{\raisebox{0.6ex}{\scriptsize{*}}} \!
      \Delta^{\mu \nu}_{21} (Q) \,
      \gamma_\mu \, S_{21} (P') \,
      \gamma_{\nu} \, ,
\end{eqnarray}
where $C_f = 4/3$ and $P' = P - Q$. In (2),
\begin{equation}
S_{21} (P') =
      - 2 \pi i \, \theta (p_0') \, (\gamma \cdot P' + M) \,
      \delta (P'^{\,2} - M^2) \, ,
\end{equation}
and $\displaystyle{\raisebox{0.6ex}{\scriptsize{*}}} \!
\Delta^{\mu \nu}_{21} (Q)$ is the (21)-component of the (HTL
resummed) effective thermal gluon propagator. For the heavy muon in
the hot electron-photon plasma, in (2) and in the following, $g$
should read $e$ and $C_f = 1$.

Throughout this Letter, we keep only the terms that yield the
leading contribution of $O(g^2 \ln g)$, which, in fact, diverges
logarithmically (cf. (8) and (9) below). It is well known
\cite{pis1,leb,bur} that the relevant terms come from the
chromomagnetic part of
$\displaystyle{\raisebox{0.6ex}{\scriptsize{*}}} \!
\Delta^{\mu \nu}_{21} (Q)$ in the soft $Q = (q_0, {\bf q})$ region,
$|Q_\mu| \leq O(gT)$;
\begin{eqnarray}
& & \left(
      \displaystyle{\raisebox{0.6ex}{\scriptsize{*}}} \!
      \Delta^{\mu \nu}_{21} (Q) \right)^{\mbox{\scriptsize{mag}}} =
      \left( \tilde{\delta}^{\mu \nu}
      - \tilde{\delta}^{\mu \rho} \, \tilde{\delta}^{\nu \sigma}
      \hat{q}_\rho \, \hat{q}_\sigma
   \right)
   \left[
      \theta(q_0) + n_B (|q_0|)
   \right]
      \nonumber \\
& & \mbox{\hspace{17.5ex}} \times
   \left[
      \, \frac{1}{Q^2 - \Pi_t (q_0 + i q_0 \epsilon, q)}
      - \mbox{c.c.} \,
   \right]
      \, ,
\end{eqnarray}
\noindent where $\tilde{\delta}_{\mu \nu} = (0, 1, 1, 1 )$,
$\hat{{\bf q}} \equiv {\bf q}/q$, $q = |{\bf q}|$, and
\lq\lq c.c.'' stands for complex conjugate. In (4), $n_B(|q_0|)$ is
the Bose distribution function and \cite{wel,bra}
\begin{eqnarray}
& & \Pi_t (q_0, q) =
      \frac{3}{2} \, m_T^2
   \left[
      \left( \frac{q_0}{q} \right)^2 + \frac{q_0 \,
      (q^2 - q_0^2)}{2 q^3} \, \ln \frac{q_0+q}{q_0-q}
   \right]
      \, , \nonumber \\
& & m_T^2 =
   \left\{ \begin{array}{ll}
      \frac{1}{9} \left( 3 + \frac{N_f}{2} \right)
      (gT)^2 & \mbox{for QCD} \\
      \frac{1}{9} \, (eT)^2 &
      \mbox{for QED \, ,}
      \end{array}
      \right.  \nonumber
\end{eqnarray}
with $N_f$ the number of quark flavors that constitutes the
quark-gluon plasma. Here we introduce the spectral density
\[
\mbox{$\large{\rho}$}_t (q_0, q)  =
      - \, \frac{1}{\pi} \, Im \,
      \frac{1}{Q^2 - \Pi_t (q_0 + i \epsilon, q)} \, .
\]
By noticing that the factor $\theta (q_0)$ in (4) can be ignored,
since $n_B(|q_0|) \simeq T/|q_0| \geq O(1/g) >> 1$, we can reduce
(4) to
\begin{equation}
\left( \displaystyle{\raisebox{0.6ex}{\scriptsize{*}}} \!
\Delta^{\mu \nu}_{21} (Q) \right)^{\mbox{\scriptsize{mag}}}
      \simeq - 2 \pi i \, (\tilde{\delta}^{\mu \nu} -
      \tilde{\delta}^{\mu \rho} \, \tilde{\delta}^{\nu \sigma}
      \hat{q}_\rho \, \hat{q}_\sigma ) \,
      \frac{T}{q_0} \, \mbox{$\large{\rho}$}_t (q_0, q) \, .
\end{equation}

Substituting (5) for
$\displaystyle{\raisebox{0.6ex}{\scriptsize{*}}} \!
\Delta^{\mu \nu}_{21} (Q)$ in (2) and using (3) with
$\delta (P'^{\, 2} - M^2) \simeq \delta ({\bf v} \cdot
{\bf q} - q_0)/(2 E)$, we obtain
\begin{eqnarray}
\Sigma_{21} (P) \simeq
      i \, g^2 \, C_f \, \frac{T}{E}
      \int \frac{d{\, ^4} Q}{(2 \pi)^2} \, \frac{1}{q_0} \,
      \delta ({\bf v} \cdot {\bf q} - q_0)\,
   \left\{
      E \gamma_0 - ( {\bf p} \cdot \hat{{\bf q}}) \,
      ( \mbox{\boldmath $\gamma$} \cdot \hat{{\bf q}} ) - M
   \right\}
      \mbox{$\large{\rho}$}_t (q_0, q) \, . \nonumber
\end{eqnarray}

It is also well known \cite{pis1,leb,bur} that the dominant
contribution to $\Sigma_{21} (P)$ or ${\cal R}_d$ comes from the
region ${\bf v} \cdot {\bf q} = q_0 << v q$, where
\[
\mbox{$\large{\rho}$}_t (q_0, q) \simeq
      \frac{3 \, m_T^2}{4}
      \frac{q \, q_0}
      {q^6 + \left( \frac{3 \pi}{4} \, m_T^2 \right)^2 q_0^2}
      \, .
\]
Using this approximation for $\mbox{$\large{\rho}$}_t (q_0, q)$, we
obtain
\begin{eqnarray}
\Sigma_{21} (P) & \simeq &
      \frac{3 \pi}{2} \, i \, g^2 \, C_f \, \frac{T}{E} \, m_T^2 \,
      (E \gamma_0 - M) \nonumber \\
& & \times
      \int \frac{d^{\, 3} q}{(2 \pi)^3} \,
      \frac{q}{q^6 + \left( \frac{3 \pi}{4} \, m_T^2 \right)^2
      ({\bf v} \cdot {\bf q})^2} \, .
\end{eqnarray}
This expression is valid for $q \leq O(g T)$ and ${\bf v} \cdot
\hat{{\bf q}} << 1$. The integral in (6) diverges
\cite{pis1,leb,bur} logarithmically at the infrared limit
$q \to 0$;
\begin{eqnarray}
\Sigma_{21} (P) & \simeq &
      \frac{i}{\pi^2} \, g^2 \, C_f \, \frac{T}{p} \,
      (E \gamma_0 - M) \nonumber \\
& & \times \int_{0^+}^{O(m_T)}
      \frac{dq}{q} \arctan \left( \frac{v m_T^2}{q^2} \right) \\
& \simeq &
      \frac{i}{2 \pi} \, g^2 \, C_f \, \frac{T}{p} \,
      (E \gamma_0 - M)
      \left[\, \ln \frac{m_T}{0^+} + O(1) \,
      \right] \, .
\end{eqnarray}
Inserting (8) into (1), we obtain
\begin{equation}
{\cal R}_d \simeq
      \frac{1}{2 \pi} \, g^2 \, C_f \, v \, T
      \left[\, \ln \frac{m_T}{0^+} + O(1) \,
      \right] \, .
\end{equation}

The damping rate, $\gamma$, is defined \cite{pis1,leb,bur} to be
$i$ times the imaginary part of the pole with $Re \, p_0 > 0$ of
the quasiparticle propagator $\{P \cdot \gamma - M -
\tilde{\Sigma}_F (P) \}^{-1}$, where $\tilde{\Sigma}_F (P)$ $=$
$- \Sigma_{22} (p_0 (1 + i \epsilon), {\bf p} )$
$- \Sigma_{21} (p_0 (1 + i \epsilon), {\bf p} )$ (cf. Chapt. 3 of
\cite{lan}) . From (8) and (12)-(14) below, we obtain, for
$p_0 \simeq E$,
\begin{eqnarray}
Im \, \Sigma_{21} (P) & \simeq & - 2 \, Im \, \Sigma_{22} (P)
      \nonumber \\
& \simeq & \frac{E}{p^2}
   \left(
      E \, \gamma_0 - M
   \right) \, {\cal R}_d \, .
      \nonumber
\end{eqnarray}
Then we see that, to the present approximation,
\begin{equation}
\gamma = \frac{1}{2} \, {\cal R}_d \, .
\end{equation}
Thus, within the HTL resummation scheme,  the damping rate $\gamma$
of the moving heavy quark is logarithmically divergent, the
well-known result which became to be an opening of a variety of
work \cite{pis1,leb,bur} under the name of \lq\lq damping-rate
puzzle''.

In an experiment, we \lq\lq detect'' the quark with momentum
${\bf p}' = {\bf p} -{\bf q}$ (Fig. 1). Then, the independent
variables are $p'$ and $Z \equiv \hat{{\bf p}} \cdot
\hat{{\bf p}}'$. Instead of using these variables, we use
$q = |{\bf p} - {\bf p}'|$ and $z \equiv \hat{{\bf p}} \cdot
\hat{{\bf q}}$; $\, p' Z = p - q z$ and $p'^2 = p^2 + q^2 - 2 p q z
\simeq p^2 - 2pqz$. As mentioned above, the dominant contribution
to ${\cal R}_d$ comes from the small region $z = q_0 / (v q)<< 1$.
Then, we consider the differential decay rate $d {\cal R}_d / dq$,
where the integration over $z$ has been carried out. As is seen in
(7), at $q \simeq 0$, ${\cal R}_d$ exhibits a $dq/q$ spectrum,
causing the logarithmic divergence. It should be recalled here that
a detector has a finite resolution with a typical resolution
$\Delta q$. Then, what we measure as $d {\cal R}_d / dq$
\rule[-2.5mm]{0.2mm}{5.7mm}
\raisebox{-2.2mm}{\scriptsize{$q < \Delta q$}} in an experiment is
\begin{eqnarray}
& & \left[
      \frac{d {\cal R}_d}{d q} \, (\Delta q)
   \right]_{q < \Delta q} \equiv
      \int_{0^+}^{\Delta q} \frac{d {\cal R}_d}{d q} \, dq
      \nonumber \\
& & \mbox{\hspace{17.8ex}} \simeq
      \frac{1}{2 \pi} \, g^2 \, C_f \, v \, T
   \left[
      \, \ln \frac{\Delta q}{0^+} + O (1) \,
   \right] \, . \nonumber \\
\end{eqnarray}
We will see below that the diverging decay rate (11) is compensated
by the analogous term in the \lq\lq elastic process''.

Now we consider the contribution of the diagrams with
\lq\lq HTL-resummed virtual'' gluon, Fig. 1 with $i = j = 1$ and
with $i = j = 2$;
\begin{eqnarray}
{\cal R}_{\lq\lq virtual \mbox{\scriptsize{''}}} & \equiv &
      - i \, \frac{M}{E} \, \frac{1}{2}
      \sum_s \bar{u}_s (P)
      [\, \Sigma_{11} (P) + \Sigma_{22} (P) \, ] u_s (P)
      \nonumber \\
& = & g^2 \frac{M}{2 E} \, C_f
      \int \frac{d{\, ^4} Q}{(2 \pi)^4}
      \sum_{\ell=1}^2  \sum_s \bar{u}_s (P)
   \left[
      \, \gamma_\mu \, S_{\ell \ell} (P') \, \gamma_\nu  \,
      \displaystyle{\raisebox{0.6ex}{\scriptsize{*}}} \!
      \Delta^{\mu \nu}_{\ell \ell} (Q) \,
   \right] u_s (P) \, ,
\end{eqnarray}
\noindent where
\[
S_{\ell \ell} (P') =
     (-)^{\ell - 1} \frac{\gamma \cdot P' + M}{P'^{\, 2} - M^2 -
     i (-)^\ell
     \epsilon} \mbox{\hspace{5ex}} (\ell = 1, 2) \, .
\]

As in the case of ${\cal R}_d$ above, the magnetic parts of
$\displaystyle{\raisebox{0.6ex}{\scriptsize{*}}} \!
\Delta^{\mu \nu}_{\ell \ell} (Q)$ $(\ell$ $=$ $1,$ $2)$ in the
region $|q_0| << q \leq O(gT)$ dominates the integral in (12), where
\begin{eqnarray}
\left( \displaystyle{\raisebox{0.6ex}{\scriptsize{*}}} \!
\Delta^{\mu \nu}_{11} (Q) \right) ^{\mbox{\scriptsize{mag}}}
& = & -
   \left\{
      \left( \displaystyle{\raisebox{0.6ex}{\scriptsize{*}}} \!
      \Delta^{\mu \nu}_{22} (Q) \right)^{\mbox{\scriptsize{mag}}}
   \right\}^{*} \nonumber \\
& = &
   \left( \tilde{\delta}^{\mu \nu}
      - \tilde{\delta}^{\mu \rho} \, \tilde{\delta}^{\nu \sigma}
      \hat{q}_\rho \, \hat{q}_\sigma
   \right) \nonumber \\
& & \times \left[
      \, \frac{1 + n_B (|q_0|)}{Q^2 -
      \Pi_t (q_0 + i q_0 \epsilon, q)} - \frac{n_B (|q_0|)}{Q^2 -
      \Pi_t (q_0 - i q_0 \epsilon, q) } \,
   \right] \nonumber \\
& \simeq &
      \left( \displaystyle{\raisebox{0.6ex}{\scriptsize{*}}} \!
      \Delta^{\mu \nu}_{21} (Q) \right)^{\mbox{\scriptsize{mag}}}
      \, .
\end{eqnarray}
\noindent Substituting (13) with (5) for
$\displaystyle{\raisebox{0.6ex}{\scriptsize{*}}} \!
\Delta^{\mu \nu}_{\ell \ell} (Q) \; (\ell = 1, 2)$ in (12) and
using $S_{11} (P') \, + \, S_{22} (P') = - 2 \pi i \,
(\gamma \cdot P' + M) \, \delta (P'^{\, 2} - M^2)$, we obtain
\begin{equation}
\Sigma_{11} (P) + \Sigma_{22} (P) \simeq - \Sigma_{21} (P) \, ,
\end{equation}
and then
\begin{equation}
{\cal R}_{\lq\lq virtual \mbox{\scriptsize{''}}} = - {\cal R}_d
\, .
\end{equation}

It is to be noted that the phys\-i\-cally mea\-sur\-able quan\-tity
\lq\lq at $q <$ $\Delta q$'', \\
\noindent $[ ( d {\cal R} / dq) \,
(\Delta q) ]_{q < \Delta q}$, is \cite{mut}
\begin{eqnarray}
\left[
      \frac{d {\cal R}}{d q} \, (\Delta q)
   \right]_{q < \Delta q} & = &
   \left[ \, 1 +  {\cal R}_{\lq\lq virtual \mbox{\scriptsize{''}}}
   \right] \nonumber \\
 & & + \left[
      \, \frac{d {\cal R}_d}{d q} \, (\Delta q) \,
   \right]_{q < \Delta q} \, .
\end{eqnarray}
The factor $1$ on the r.h.s. of (16) is the zeroth-order
contribution. Substituting (11) and (15) with (9) into (16), we see
that the cancellation of divergences occurs between
\lq\lq virtual''- and \lq\lq real''-contributions, and find
\[
\left[
      \frac{d {\cal R}}{d q} \, (\Delta q)
\right]_{q < \Delta q} \simeq
      1 - \frac{1}{2 \pi} \, g^2 \, C_f \, v \, T
   \left[
      \, \ln \frac{m_T}{\Delta q} + O(1) \,
   \right] \, .
\]

Thus the measurable quantity, although sensitive to the resolution
$\Delta q$, is free from divergence. The mechanism of cancelling
divergences in (16) is exactly the same as in the cancellation of
infrared divergences in vacuum theory (Bloch-Nordsieck mechanism)
\cite{mut}. In some reaction rate in QED at $T = 0$, both the
diagram with electron (soft) bremsstrahlung and the diagram with
radiative correction diverge, but they cancel each other: The
former is the counterpart to
$[ \, (d {\cal R}_d / d q) \, (\Delta q) \, ]_{q < \Delta q}$
(Eq. (11)), while the latter is the counterpart to
${\cal R}_{\lq\lq virtual \mbox{\scriptsize{''}}}$ (Eq. (15)). It
is worth pointing out, in passing, that we obtain
\[
\left[ \frac{d {\cal R}}{d q}\, (\Delta q) \right]_{q < \Delta q} +
      \int_{\Delta q}^{O (m_T)} \frac{d {\cal R}_d}{d q} \, d q = 1
      \, ,
\]
as it should be.

Now we are in a position to discuss the physical content of (16) in
terms of the processes taking place in the quark-gluon plasma. For
any given diagram in the real-time formalism, general rules of
identifying the physical processes are available \cite{nie}: Given
a (thermal) diagram, like Fig. 1, representing some thermal
reaction rate, one  may divide it into two parts; the one is the
reaction's $S$-matrix element in {\em vacuum theory} and the other
is the $S^*$-matrix element. The $S$- and $S^*$-matrix elements
represent the reactions between the considered particle(s) (the
heavy quark in the present case) and the particles in the
quark-gluon plasma. The type-1 (type-2) vertices in the thermal
diagram go to the vertices in the $S$- ($S^*$-) matrix element. A
thermal propagator with momentum $K$ from a type-1 vertex to a
type-2 vertex is \lq\lq cut''. When $k_0 > 0$ ($k_0 < 0$) the
\lq\lq cut'' is the \lq\lq final-state cut''
(\lq\lq initial-state cut''). For a thermal propagator from a
type-2 vertex to a type-1 vertex, the opposite rules apply. For
more details, we refer to \cite{nie}.

We note that $\displaystyle{\raisebox{0.6ex}{\scriptsize{*}}} \!
\Delta^{\mu \nu}_{21}$ in (2) and
$\displaystyle{\raisebox{0.6ex}{\scriptsize{*}}} \!
\Delta^{\mu \nu}_{\ell \ell}$ in (12) may be expanded in powers of
$\Pi_t$ (cf. (4) and (13)), and thus ${\cal R}_d$ and
${\cal R}_{\lq\lq virtual \mbox{\scriptsize{''}}}$ turn out to be
represented by infinite series; ${\cal R}_d = \sum_{i = 0}^\infty
{\cal R}_d^{(i)}$ and
${\cal R}_{\lq\lq virtual \mbox{\scriptsize{''}}} =
\sum_{i = 0}^\infty
{\cal R}_{\lq\lq virtual \mbox{\scriptsize{''}}}^{(i)}$. From each
contribution thus obtained, we can identify the physical processes
according to the rules outlined above. An example of the physical
process that is involved in ${\cal R}_d^{(2)}$
(${\cal R}_{\lq\lq virtual \mbox{\scriptsize{''}}}^{(2)}$) is
depicted in Fig. 2 with the final-state cut line $C_d$ ($C_v$); the
left side part of the cut line represents the $S$-matrix element in
{\em vacuum theory}, while the right part represents the
$S^{*}$-matrix element. The group of particles on top of Fig. 2
stands for spectators which are constituent particles (quarks,
antiquarks and gluons) of the quark-gluon plasma. In Fig. 2, $Q$ is
soft $\sim O(gT)$, while $K_1, \, K_2, \, P_1$ and $P_2$ are hard
$\sim O(T)$. It is to be noted that, in Fig. 2 with the final-state
cut line $C_v$, the heavy quark in the $S^*$-matrix element is
simply a spectator particle.

For $Q \simeq 0$, $P \simeq P'$, and then the \lq\lq heavy quark
detector'' cannot discriminate the final-state heavy quark ($P'$)
in Fig. 2 with the final-state cut line $C_d$ and the one ($P$) in
Fig. 2 with $C_v$. Then, what we measure corresponds to the sum of
Fig. 2 with $C_d$ and Fig. 2 with $C_v$, as well as of all other
diagrams included in ${\cal R}_d$, Eq. (1), and
${\cal R}_{\lq\lq virtual \mbox{\scriptsize{''}}}$, Eq. (12), (cf.
(16)).

We like to emphasize again that two heavy quarks, one with $P' = P
\, \, (q = 0)$ and another with $P' \, \, (|{\bf p}' - {\bf p}| <
\Delta q)$, cannot be discriminated experimentally. In other words,
one cannot recognize the heavy quark with $|{\bf p}' - {\bf p}| <
\Delta q$ as a (thermal) decay product, which is responsible for
diverging damping rate $\gamma$, Eq. (10). On the basis of this
observation, we propose to introduce a \lq\lq observable damping
rate'' defined as (cf. (10))
\begin{eqnarray}
\gamma_{\mbox{\scriptsize{obs}}} (\Delta q) & = & \frac{1}{2}
   \left\{
      {\cal R}_d -
   \left[
      \frac{d {\cal R}_d}{d q}\, (\Delta q)
   \right]_{q < \Delta q}
   \right\} \nonumber \\
& \simeq & \frac{1}{4 \pi} \, g^2 \, C_f \, v \, T \,
      \ln \left( \frac{m_T}{\Delta q} \right) \, .
      \nonumber
\end{eqnarray}
This $\gamma_{\mbox{\scriptsize{obs}}} (\Delta q)$ is the damping
rate that originates from the heavy-quark decay into physically
distinguishable states.

It is to be noted that, for a heavy quark at rest, the damping rate
$\gamma$ is infrared-safe and finite quantity of $O( g^2 T)$
\cite{pis1,bra1}, and then the difference between $\gamma$ and
$\gamma_{\mbox{\scriptsize{obs}}} (\Delta q)$ is negligibly small.
The reason for the infrared safety of $\gamma$ is traced back to
the fact that the heavy quark at rest couples only to the
chromoelectric parts of
$\displaystyle{\raisebox{0.6ex}{\scriptsize{*}}} \!
\Delta^{\mu \nu}_{21} (Q)$ and
$\displaystyle{\raisebox{0.6ex}{\scriptsize{*}}} \!
\Delta^{\mu \nu}_{\ell \ell} (Q) \, \, (\ell = 1, 2)$, which, in
contrast to the chromomagnetic part, develops a (thermal) Debye
mass and there emerges the energy gap between the single
heavy-quark state and the states with a heavy quark plus a
\lq\lq HTL-resummed gluon'' with $Q \simeq 0$.


\newpage
\begin{center}
{\Large {\bf Figure captions} }\vspace*{1.5em} \\
\end{center}

\noindent Fig.1. A ther\-mal self-energy di\-a\-gram of a heavy
quark in real-time thermal field \\
\hspace*{7ex} theory. $i$ and $j$ ($i,$ $j$ $ =$
$1,$ $2$) designate the type of vertex. The blob indicates \\
\hspace*{7ex} the (HTL resummed) effective gluon propagator.

\noindent Fig.2. Typical processes taking place in the quark-gluon
plasma. The diagram with \\
\hspace*{7ex} the final-state cut line $C_d$ ($C_v$) is
the process that is included in ${\cal R}_d$
(${\cal R}_{\lq\lq virtual \mbox{\scriptsize{''}}}$) \\
\hspace*{7ex} or in Fig. 1
with $i = 1$ and $j = 2$ ($i = 1$ and $j = 1$). The left side of
the \\
\hspace*{7ex} final-state cut line represents the $S$-matrix
element, while
the right side  rep- \\
\hspace*{7ex} resents the $S^*$-matrix element in vacuum
theory.

\end{document}